# Adoption and Ecosystem Health: A Longitudinal Analysis of Open-Source Multi-Agent Frameworks

**Xi Zhang, PhD** **Cisco Systems**

**Papi Menon** **Cisco Systems**

**Vivian Chu** **Cisco Systems**

**Koray Cosguner, PhD** **Indiana University**

## Abstract

Since ChatGPT's launch in November 2022, open-source agentic frameworks have proliferated, making framework selection important for engineering teams while obscured by popularity signals such as GitHub stars. This paper analyzes 15 major open-source AI agent framework repositories from late 2022 to early 2026, using 808,042 stars, 73,997 pull requests, 86,241 commits, and 987,330 user profiles to assess ecosystem health across awareness, adoption, and retention. Three findings emerge. First, headline popularity is unreliable. Star counts reflect hype cycles and inorganic activity. AutoGPT gained 111,967 stars in one month but converted fewer than 9 contributors per 1,000 stars, defined as contributor density in this research, compared with LangChain's 41. Lower-profile frameworks such as Pydantic-AI show higher contributor density, indicating deeper adoption. Second, mapping awareness against adoption shows that visibility and engagement diverge. MetaGPT and LangFlow have contributor density ratios below 5 even with their high visibility. Openai-agents-python's limited contributor base suggests institutional backing alone does not ensure community depth. By analyzing cross-framework contribution, we discover that LangChain functions as a shared infrastructure, attracting 82.5% of cross-ecosystem contributors. Third, retention drops most steeply in the first 30 days of initial contribution and stabilizes near 90 days. Overall, ecosystem health is better measured by contributor density, cross-ecosystem engagement, and retention than by stars alone. These metrics offer teams a more robust basis for framework evaluation.

## 1. Introduction

The evolution of agentic AI frameworks reflects a broader industry understanding of deploying reliable AI in real-world settings, beyond mere software adoption [1]. Following ChatGPT's public launch in late 2022 [2], the 2023–2024 period signaled the rise of autonomous agents, with frameworks like LangChain enabling composable LLM pipelines and AutoGPT pioneering autonomous agent loops for reasoning and acting [3]. Each wave of new frameworks directly addresses the shortcomings of previous versions, shifting from unrestricted autonomy to modular chaining, from single-agent to multi-agent systems [4], and from experimental prototypes to type-safe, observable, enterprise-ready solutions. As public interest in open-source agentic AI continues to grow, an increasing number of enterprise AI strategies now incorporate open-source frameworks as a core architectural component, making framework selection a consequential and understudied decision. Betting on the wrong ecosystem creates business risk, including inherited technical debt, a shrinking contributor base, and reduced support if a corporate backer changes direction.

Many metrics used to evaluate open-source AI projects are often misunderstood and susceptible to manipulation. For example, star counts tend to indicate popularity rather than actual adoption [5], while contributor numbers can include a single person's quick bug fix alongside years of ongoing effort [6]. The volume of pull requests may also be artificially increased through automated updates or AI-generated patches. Although these metrics might suggest rapid growth and a healthy ecosystem, closer examination reveals that sustained adoption varies considerably across frameworks and appears strongly associated with underlying architectural choices. Frameworks that achieve long-term, production-level adoption tend to address specific, tangible engineering challenges, suggesting that raw growth metrics alone are insufficient predictors of staying power. To examine this, the paper proceeds in three analytical layers: first, we assess community awareness through GitHub star trajectories; second, we analyze adoption depth through contributor patterns across ecosystems; third, we evaluate sustainability through contributor retention over time.

## 2. Data

This study draws on publicly available data from GitHub, collected via the GitHub REST API and GraphQL API. The dataset spans multiple dimensions of repository activities, including star events, commit histories, pull request records, and user profile, enabling analysis across three analytical layers: community awareness, active adoption, and contributor retention. All data was collected from public repositories and user profiles. No private or authenticated user data was accessed.

### 2.1 Repository Selection

The following 15 repositories were selected for analysis based on their prominence within the open-source AI framework and orchestration ecosystem, the distinct role each plays in its development, and the availability of sufficient public activity to support comparative analysis. This selection introduces a degree of survival bias: repositories that remained visible and active through early 2026 are more likely to be included than projects that were short-lived, archived early, or failed to gain traction. As a result, the dataset should be interpreted as a study of leading and enduring open-source agent frameworks rather than a complete census of all frameworks launched during the period.

| Repo | Organization | Repo Created | Primary Use Case |
|---|---|---|---|
| **langchain** | langchain-ai | Oct, 2022 | General-purpose LLM application framework |
| **langflow** | langflow-ai | Feb, 2023 | Low-code visual RAG and agent builder |
| **semantic-kernel** | microsoft | Feb, 2023 | Enterprise LLM SDK for .NET and Python |
| **AutoGPT** | Significant-Gravitas | Mar, 2023 | Pioneer autonomous agent platform |
| **MetaGPT** | FoundationAgents | Jun, 2023 | Multi-agent framework with SOP-based role assignments [7] |
| **langgraph** | langchain-ai | Aug, 2023 | Stateful multi-actor graph orchestration |
| **autogen** | microsoft | Aug, 2023 | Conversational multi-agent framework [8] |
| **crewAI** | crewAIInc | Oct, 2023 | Role-based multi-agent task orchestration |
| **agentscope** | agentscope-ai | Jan, 2024 | Multi-agent platform with fault tolerance |
| **pydantic-ai** | pydantic | Jun, 2024 | Type-safe agent framework built on Pydantic |
| **mastra** | mastra-ai | Aug, 2024 | TypeScript-native agent and workflow framework |
| **smolagents** | huggingface | Dec, 2024 | Lightweight code-first agent framework |
| **openai-agents-python** | openai | Mar, 2025 | First-party OpenAI Python agent SDK |
| **adk-python** | google | Apr, 2025 | Google-backed agent development toolkit |
| **agent-framework** | microsoft | Apr, 2025 | Microsoft unified agent framework (successor to SK + AutoGen) |

Note: GitHub repository URLs for all 15 frameworks analyzed in this study are as follows:
LangChain (https://github.com/langchain-ai/langchain),
Langflow (https://github.com/langflow-ai/langflow),
Semantic Kernel (https://github.com/microsoft/semantic-kernel),
AutoGPT (https://github.com/Significant-Gravitas/AutoGPT),
MetaGPT (https://github.com/FoundationAgents/MetaGPT),
LangGraph (https://github.com/langchain-ai/langgraph),
AutoGen (https://github.com/microsoft/autogen),
CrewAI (https://github.com/crewAIInc/crewAI),
AgentScope (https://github.com/agentscope-ai/agentscope),
Pydantic AI (https://github.com/pydantic/pydantic-ai),
Mastra (https://github.com/mastra-ai/mastra),

smolagents (https://github.com/huggingface/smolagents),
OpenAI Agents (https://github.com/openai/openai-agents-python),
Google ADK (https://github.com/google/adk-python), and
Agent Framework (https://github.com/microsoft/agent-framework).
All repositories were accessed on March 10, 2026.

### 2.2 Data Collected

Data was collected from the GitHub API across four event types for each of the 15 repositories: stars, pull requests, commits, and user profiles. Collection covered each repository's full public history from its creation date through March 2026, spanning repositories with a creation window from October 2022 to May 2025. In aggregate, the dataset contains 808,042 stars, 73,997 pull requests, 86,241 commits, and 987,330 globally deduplicated user profiles. Each profile record includes account creation date, company affiliation, and biographical fields, which are used to compute account age, assess profile completeness, and identify organization-affiliated contributors in the subsequent analyses.

Throughout the analyses, contributors are restricted to authors of pull requests and commits. We exclude issue authors because they do not reflect direct code contribution. Bot accounts are excluded from contributor counts, identified through username pattern matching against known automation account conventions including 'bot,' 'dependabot,' 'github-actions,' 'copilot,' 'renovate,' 'mergify,' 'stale,' 'codecov,' and 'snyk.' This heuristic may undercount bot exclusions where non-standard naming conventions are used. After filtering, the dataset contains 12,594 unique human code contributors across all 15 repositories.

### 2.3 Analytical Methods

This study analyzes repository activities across three dimensions: community awareness, adoption depth, and contributor retention.

Community awareness is operationalized as total GitHub stars, treated as a directional indicator of visibility rather than a definitive adoption measure. To assess the reliability of star-based metrics, each starring event is evaluated against three signals: new account age, empty profile, and temporal burst. Events triggering two or more signals are flagged as anomalous [9].

Adoption depth is operationalized as total code contributors, restricted to pull request and commit authors, with a contributor density ratio, defined as total contributors per 1,000 stars, used to normalize comparison across repositories with different visibility levels. Contributor profiles are examined across two dimensions: median GitHub account age and company affiliation, both derived from self-reported GitHub profile fields. We further examine cross-ecosystem adoption by identifying contributors who crossed organizations' boundaries, following prior work showing that such patterns reflect meaningful signals of framework adoption and interoperability.

Furthermore, we jointly visualize awareness and adoption in a quadrant framework segmenting repositories into Market Leaders, Quiet Compounders, Momentum Traps, and Nascent Entrants, with awareness operationalized as total GitHub stars and adoption as total code contributors.

Finally, we measure contributor retention using a cumulative cohort approach focused on early contributors, defined as those who made their first contribution within the first 180 days of a repository's creation. Retention is measured at 30-day intervals up to 360 days, with each interval recording the cumulative proportion of contributors who returned to make at least one additional pull request or commit by that checkpoint. This approach distinguishes frameworks that build lasting contributor relationships from those that attract predominantly one-time participants, while focusing on the early cohort ensures that retention behavior remains comparable across repositories with different launch dates and cohort structures.

## 3. Awareness

The primary gauge of public awareness for an open-source project is its GitHub star count. Across the 15 projects that were studied in this research, star growth varied considerably (see Figure 1 for cumulative star trajectories plotted against calendar time and Appendix A for individual star histories showing both cumulative accumulation and monthly rate of change for each repository).

*Figure 1: Cumulative GitHub Stars Over Time Across 15 Open-Source AI Frameworks*

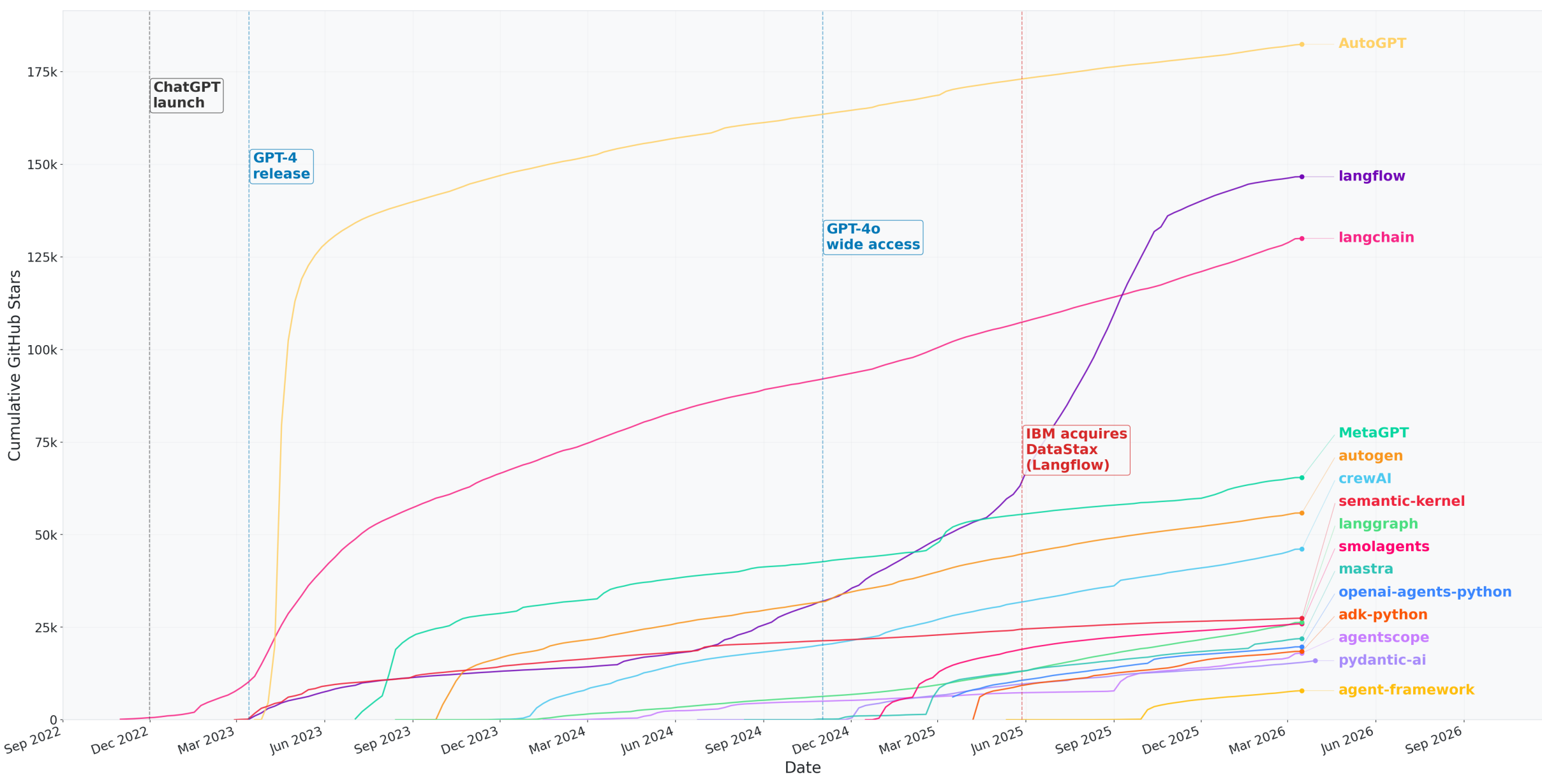

### 3.1 Cumulative Star Trajectories: A Cross-Framework Overview

AutoGPT's surge in April 2023 marked a significant inflection point in the early ecosystem, accumulating 111,967 stars in a single month and surpassing all other frameworks in this dataset. This spike was largely driven by widely circulated social media demonstrations in which AutoGPT purportedly conducted research, wrote code, and planned travel autonomously. However, growth declined sharply within a few months, demonstrating a pattern of hype-driven interest that failed to translate into lasting community engagement. One contributing factor may have been AutoGPT's tendency to enter recursive loops during execution, draining API credits and diminishing practical utility once initial enthusiasm subsided.

By comparison, LangChain and Langflow demonstrate more sustained growth in star accumulation. LangChain's trajectory is notably steady, averaging approximately 3,000 stars per month over 41 months, suggesting consistent demand-driven adoption rather than a viral spike. This pattern is consistent with ongoing developer discovery through tutorials, blog posts, and enterprise AI onboarding materials, indicative of a robust ecosystem effect rather than a single promotional event. Langflow, though launched later, has surpassed LangChain in total stars (146k vs. 130k as of March 10, 2026). Its peak of 16,388 stars in September 2025 likely reflects sustained enterprise attention following IBM's announced acquisition of DataStax, Langflow's parent company, in February 2025 [10], which positioned Langflow as part of IBM's watsonx AI portfolio and significantly raised its enterprise visibility throughout the year.

While Figure 1 highlights the impact of market timing, Figure 2 normalizes trajectories to months since launch, standardizing the observation window to compare each framework's adoption velocity across its lifecycle.

*Figure 2: Star Accumulation Chart Normalized to Months Since Launch*

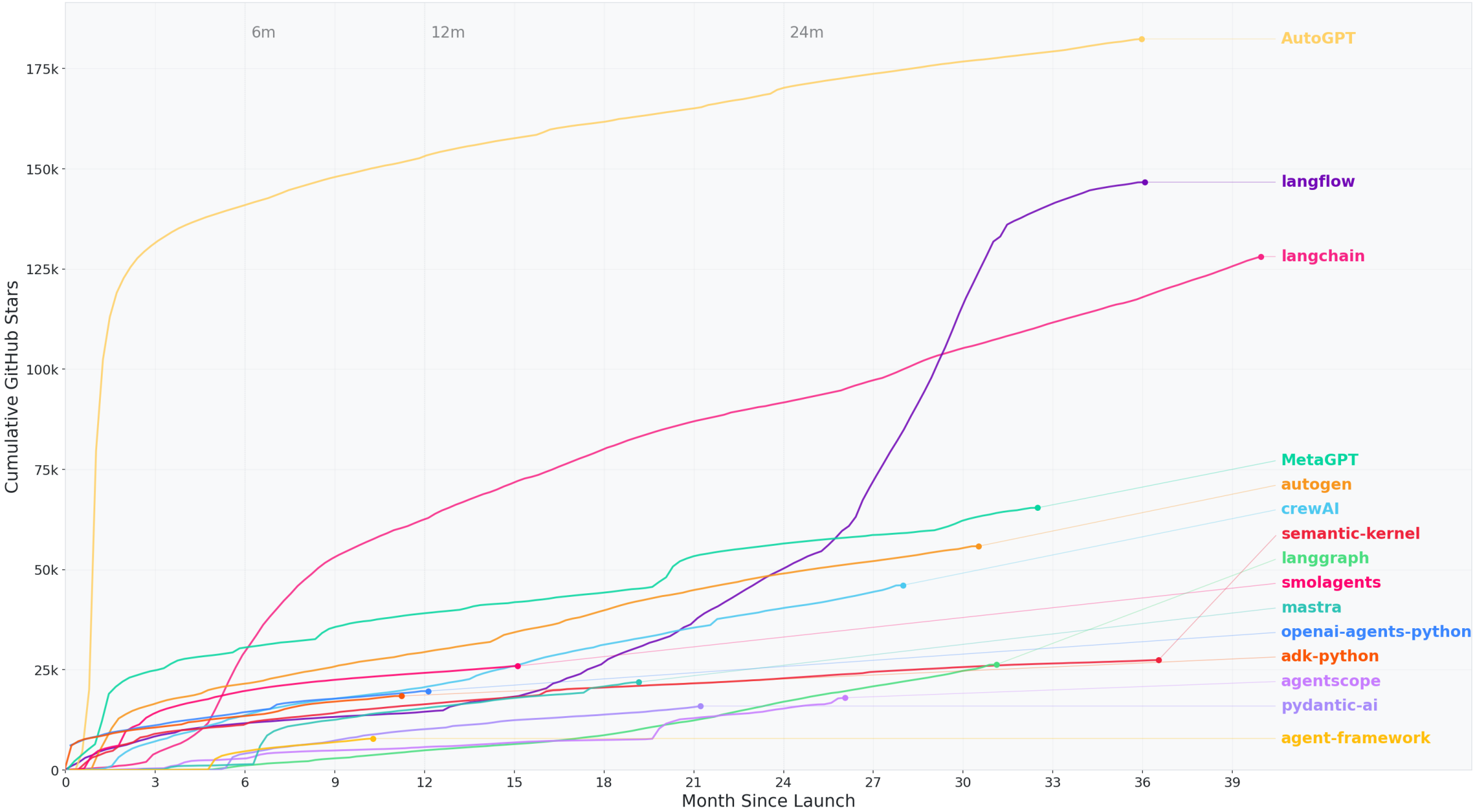

The predominant pattern across our dataset is the classic S-curve lifecycle, in which rapid initial adoption gradually decelerates as community interest stabilizes. In our data, growth appears to plateau early. By month 24, most repositories were expanding at less than half their initial rate.

The second pattern is late acceleration. Langflow ranked 10th among the 15 frameworks at six months post-launch but climbed to 2nd place by month 30. Its lifecycle began with modest initial traction before a sharp inflection in its third year, when star count rose from 48,680 to 146,224, approximately a 200% increase.

The third pattern is steady compounding. LangChain's star count grew from 30,848 at month 6 to 118,477 at month 36, nearly quadrupling over that period. This trajectory suggests that sustained adoption reflects genuine utility and ongoing ecosystem value rather than any single promotional event.

## 3.2 The Institutionalization Phase: First-Party SDKs

The latest group in the dataset, including openai/openai-agents-python, google/adk-python, and huggingface/smolagents, marks a new stage characterized by official frameworks released directly by foundation model ecosystem leaders. Their emergence indicates that the agentic AI ecosystem has matured sufficiently to warrant direct first-party development investment.

*Figure 3: Cumulative Stars for First Party SDKs*

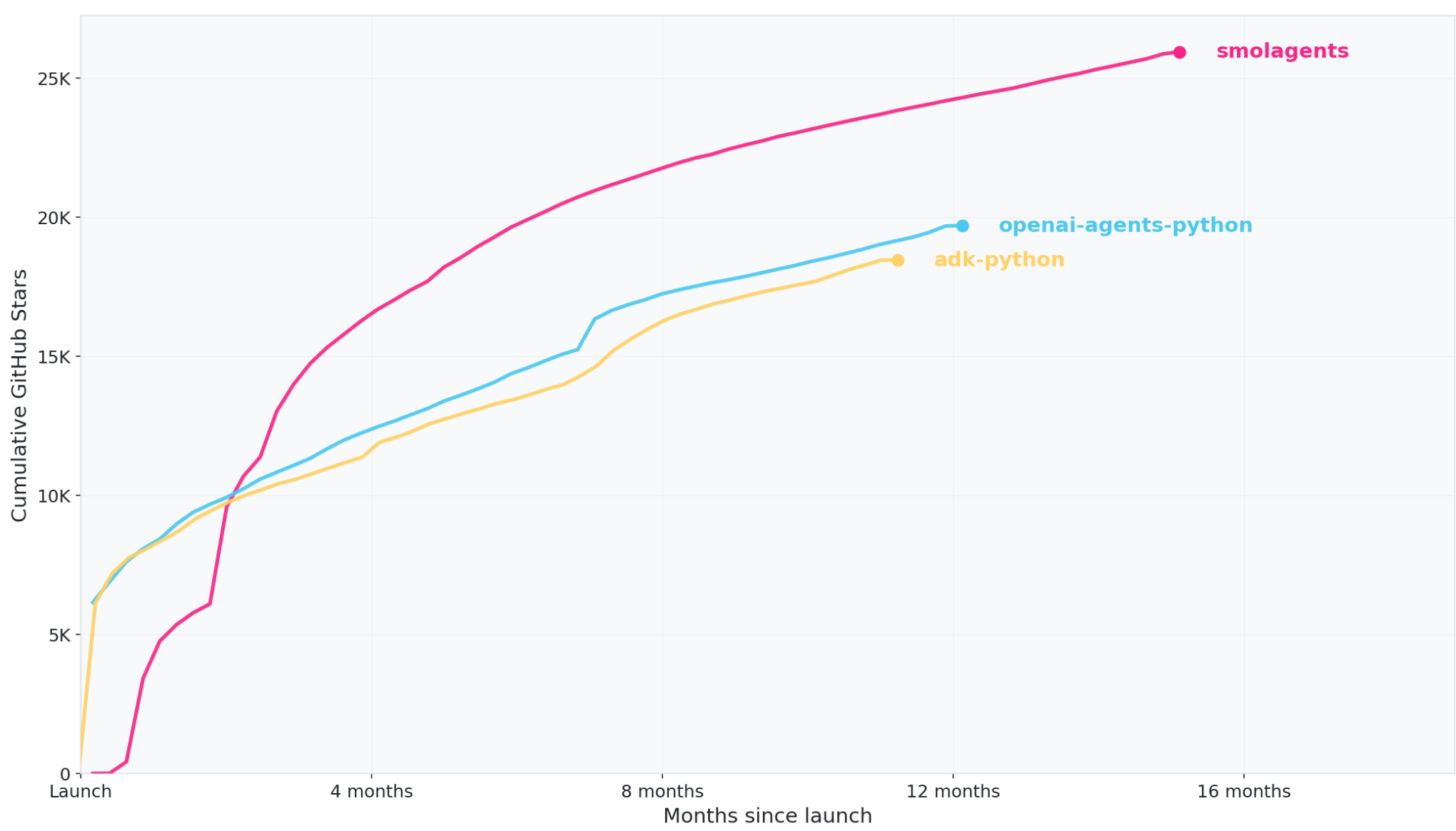

As shown in Figure 3, their initial star counts are notable but not exceptional within the broader dataset (compared to, for example, LangChain's 29,012 stars at month 6). Smolagents reached 20,469 stars by month 6 and 24,553 by month 12, ranking fifth at month 6. OpenAI Agents SDK achieved 15,080 stars at month 6 and 19,699 at month 12. Google ADK recorded 14,204 stars at month 6, with no data available afterward because the repository had not yet reached month 12 at the time of data collection. The three repositories cluster between 14,000 and 21,000 stars at month 6, indicating that institutional origin provides a significant, though not dominant, advantage at launch.

Whether this early traction translates into sustained compound growth remains an open empirical question. The deceleration is already visible in smolagents between months 6 and 12 with a 20% gain, representing the lowest first-interval growth rate in the dataset. It warrants attention though additional milestones beyond month 18 will be required for a definitive characterization. Those subsequent data points will serve as critical tests of whether institutional support fosters enduring community engagement or merely accelerates initial discovery.

### 3.3 Anomalous Star Patterns

The high star counts observed in some repositories prompted an examination of whether these growth patterns reflect typical developer interest or unusually concentrated starring behavior. A large-scale study of GitHub starring activity identified six million suspected fake stars between 2019 and 2024 [9], with AI and LLM repositories representing the largest non-malicious category of recipients.

To investigate this, we developed a three-signal detection framework (Figure 4), classifying any starring event that triggers two or more signals as flagged as anomalous. We choose a threshold of two or more signals to balance detection sensitivity against false positives. Any single signal in isolation may characterize legitimate users such as a new developer with an incomplete profile who starred a repository shortly after account creation.

*Figure 4: Anomalous Star Signals*

| **Signal** | **Threshold** | **Rationale** |
|---|---|---|
| New account | Account ≤ 7 days old when starring | Accounts may be created solely to star repositories |
| Empty profile | 0 of 5 profile fields filled | Bot or throwaway accounts often leave name, bio, location, email, and website blank. |
| Temporal burst | ≥ 50 stars to the same repo in the same clock hour | Organic interest usually does not sustain 50+ stars per hour; sharp bursts can indicate coordinated campaigns. |

Figure 5 highlights the three repositories with the highest percentage of flagged anomalous stars. A display threshold of 5% was applied, as rates below this level are consistent with the baseline noise level of normal GitHub activity. A baseline repository deletion ratio of 5.03% has been reported for accounts with no detected anomalous behavior [9], suggesting that repositories falling below this threshold are unlikely to reflect coordinated inorganic activity.

*Figure 5: Top Three Repositories with Flagged Stars*

| **Repo Name** | **Total Stars** | **Anomalous Stars Flagged** | **Starred by New Acct** | **Starred by Empty Profile** | **Starred during Star Burst Hour** | **Maximum Stars per Hour** | **Flagged %** |
|---|---|---|---|---|---|---|---|
| langflow-ai/langflow | 146655 | 40547 | 39944 | 82335 | 2817 | 117 | 27.7 |
| Significant-Gravitas/AutoGPT | 182405 | 30655 | 4655 | 50399 | 110803 | 1231 | 16.8 |
| openai/openai-agents-python | 19699 | 1154 | 257 | 4979 | 4624 | 228 | 5.9 |

Figure 5 illustrates clear patterns of irregular activity in the top three repositories.

- Langflow's highest spike was relatively small with 117 stars per hour, but most flagged stars originated from accounts created on the same day they starred. This pattern is consistent with elevated attention around the announced DataStax and IBM acquisition activity, though further validation would be needed to determine whether it reflects coordinated amplification, heightened organic interest, or both.

- AutoGPT shows a burst-heavy anomaly profile: 60.7% of flagged stars appeared during surge periods, with the highest spike in April 2023 amid the viral hype cycle surrounding ChatGPT's widespread adoption. This burst pattern aligns with the popularity of AutoGPT's ReAct-loop demo. One possible explanation is that the rise in genuine public interest was accompanied by unusually concentrated starring behavior.

- OpenAI Agents SDK has a shorter observation window than the other repositories. Its flagged stars are concentrated near the project's launch period, which is consistent with launch-amplified activity rather than a sustained anomalous pattern. Given the repository's age, this pattern should be interpreted cautiously.

Differentiating between coordinated artificial amplification and authentic viral engagement poses significant methodological challenges, and this analysis adopts a deliberately conservative stance. It identifies anomalous star behavior rather than verified aritificial manipulation. Stars may be acquired through star-farming services coordinated by automated bots or generated by authentic viral trends that exhibit burst-like patterns. In some cases, the extent of irregular activity may simply reflect the exceptionally high level of interest these initiatives attract as artificial intelligence development becomes increasingly accessible.

## 4. Adoption from Contributors

While star counts measure community awareness, a stronger signal of genuine adoption is active contribution through merged pull requests or commits. Contributor activity has been identified as a reliable proxy for project adoption in large-scale analyses of open-source communities [11].

### 4.1 Contributor Profile

Across the 15 repositories, contributor counts range from 145 for AgentScope to 5,362 for LangChain, with a dataset median of 618. Because raw counts alone provide limited context, we also examine two contributor profile dimensions. The median GitHub account age is 8.3 years, suggesting an experienced developer base rather than a primarily newcomer audience. Company affiliation, based on self-reported GitHub profiles, averages 38.8% but varies across repositories, from 26.4% for AgentScope to 52.5% for Agent Framework. This variation likely reflects differences in target audience, with Microsoft-family repositories consistently attracting a more professionally affiliated contributor base.

*Figure 6: Repo Contributors and Profile*

| Repo Name | Total Contributors | Median Account Age (yrs) | % Company Affiliated |
|---|---|---|---|
| langchain | 5362 | 8.9 | 37.7 |
| langgraph | 607 | 7.5 | 38.4 |
| langflow | 707 | 7.9 | 39.5 |
| MetaGPT | 257 | 7 | 30.6 |
| autogen | 869 | 8.8 | 39.7 |
| crewAI | 645 | 8.4 | 38.9 |
| semantic-kernel | 618 | 10.1 | 49.2 |
| agentscope | 145 | 7 | 26.4 |
| smolagents | 395 | 8.4 | 35.9 |
| mastra | 484 | 9.2 | 44.5 |
| openai-agents-python | 438 | 6.3 | 36.9 |
| adk-python | 621 | 7.6 | 39.2 |
| pydantic-ai | 675 | 9 | 44.5 |
| agent-framework | 204 | 9.4 | 52.5 |
| AutoGPT | 1716 | 8.3 | 28.1 |

## 4.2 Cross-Ecosystem Contribution Patterns

In addition, we also examine contributors who engage across multiple ecosystems, focusing specifically on those who crossed organizational boundaries. This methodology is grounded in prior work [12] showing that cross-ecosystem contribution patterns reflect meaningful signals of framework adoption and interoperability.

Figure 7 illustrates that LangChain functions as a key connector within the ecosystem. Of the 12,600 unique contributors across 15 frameworks, 590 users contributed across distinct organizational boundaries. LangChain appears in 11 of the top 20 cross-ecosystem contributor pairings, with approximately 82.5% of cross-ecosystem contributors having engaged with it.

This bridge pattern has a clear technical basis: LangChain's composable primitives are designed for extension rather than replacement [13]. CrewAI was originally built on LangChain's core, and LangFlow's visual interface integrates LangChain components directly. Contributors working on these downstream frameworks are also engaging with LangChain's abstraction layer. Between 68% and 87% of contributors across Langflow, CrewAI, and Pydantic AI have also contributed to LangChain. We find that these overlapping contributors are generally more experienced, with older accounts averaging 9 years and a higher PR merge rate of 73.8%. This pattern suggests that framework selection among experienced contributors reflects deliberate architectural reasoning rather than opportunistic or trend-driven adoption.

*Figure 7: Top 20 Cross-Ecosystem Contributor Combinations*

| Cross-Ecosystem Combination | # of Cross-Ecosystem Contributors | % of Total |
|---|---|---|
| LangChain + Pydantic AI | 48 | 8.1 |
| LangChain + Langflow | 48 | 8.1 |
| LangChain + AutoGen | 45 | 7.6 |
| LangChain + AutoGPT | 41 | 6.9 |
| LangChain + CrewAI | 38 | 6.4 |
| LangChain + Semantic Kernel | 18 | 3.1 |
| LangChain + Google ADK | 17 | 2.9 |
| LangChain + smolagents | 16 | 2.7 |
| LangChain + OpenAI Agents | 13 | 2.2 |
| AutoGen + CrewAI | 12 | 2 |
| AutoGen + LangGraph | 9 | 1.5 |
| LangChain + MetaGPT | 9 | 1.5 |
| LangGraph + CrewAI | 7 | 1.2 |
| LangChain + Mastra | 7 | 1.2 |
| AutoGPT + AutoGen | 7 | 1.2 |
| AutoGen + Pydantic AI | 6 | 1 |
| AutoGen + Google ADK | 6 | 1 |
| Pydantic AI + smolagents | 6 | 1 |
| LangChain + AgentScope | 5 | 0.8 |
| Google ADK + OpenAI Agents | 5 | 0.8 |

*Note: Contributors who are active only in the same organization, such as Semantic Kernel and AutoGen under Microsoft, are not counted as cross-ecosystem contributors.*

Pydantic AI and Google ADK emerge as 'hidden gems' within the ecosystem, defined as frameworks with relatively modest star counts (15,950 and 18,460 respectively, compared to a dataset median of 26,295) yet disproportionately high cross-ecosystem contributor density. This pattern likely reflects their appeal to experienced practitioners who prioritize production reliability over ecosystem popularity. Pydantic AI's emphasis on type safety, validation, and structured outputs [14] and Google ADK's design around enterprise software engineering principles [15], including modularity, testability, and deployment reliability, make both frameworks particularly attractive to the kind of experienced contributors whose elevated account ages and PR merge rates were observed across cross-ecosystem pairings in this dataset.

### 4.3 Awareness and Adoption Quadrant Analysis

While star counts reveal community awareness and cross-ecosystem contribution patterns reveal the depth of practitioner engagement, comparing total stars directly with total code contributors exposes a more nuanced picture of how frameworks build or fail to build genuine communities. Figure 8 positions the fifteen repositories into four quadrants: **high awareness/high adoption (Market Leaders)**, **low awareness/high adoption (Quiet Compounders)**, **high awareness/low adoption (Momentum Traps)**, and **low awareness/low adoption (Nascent Entrants)**.

In this research, we define awareness as total GitHub stars (x-axis, log scale), which reflect community visibility and discoverability, and adoption as total code contributors through pull requests and commits (y-axis, log scale), which reflect active engagement with the codebase. Log scales on both axes normalize the wide dispersion in repository scale, allowing newer repositories to remain visible alongside more established and high-starred projects. The quadrant dividers are set at the dataset median for both contributors and stars.

These proxies carry inherent limitations. Star counts may reflect viral attention or coordinated promotion, as shown in the anomalous stars pattern analysis. Contributor counts provide a stronger signal of sustained engagement but capture only code contributions, excluding other adoption indicators such as package downloads, API usage, or production deployments. Accordingly, both metrics should be interpreted as directional indicators rather than definitive measures of community health.

Figure 8 shows clear differences in how frameworks convert visibility into contributor participation. LangChain anchors the Market Leaders quadrant, combining the largest contributor base in the dataset with one of the highest star counts. AutoGPT also appears in this quadrant, but with a very different profile: despite having the highest star count, its contributor base is much smaller relative to its visibility. LangFlow also accumulates stars at scale but falls closer to the boundary on contributor counts, suggesting weaker community conversion relative to its visibility. LangGraph sits near the quadrant center, with moderate star counts and contributor depth relative to the other Market Leaders. This positioning may reflect its complementary role within the LangChain ecosystem rather than independent community formation.

The Quiet Compounders quadrant highlights frameworks with modest awareness but strong contributor depth. Pydantic-AI and Google ADK fall into this group, with star counts below the dataset median but contributor counts approaching the quadrant boundary. Both frameworks are relatively young, as they were launched in June 2024 and April 2025 respectively, and their proximity to the boundary on contributor counts likely reflects early-stage community growth rather than an established pattern of strong contributor engagement. Our analysis shows that Pydantic-AI also appears to attract a more experienced and company-affiliated contributor base, pointing to stronger production-oriented engagement.

*Figure 8: Ecosystem Positioning of AI Agent Frameworks*

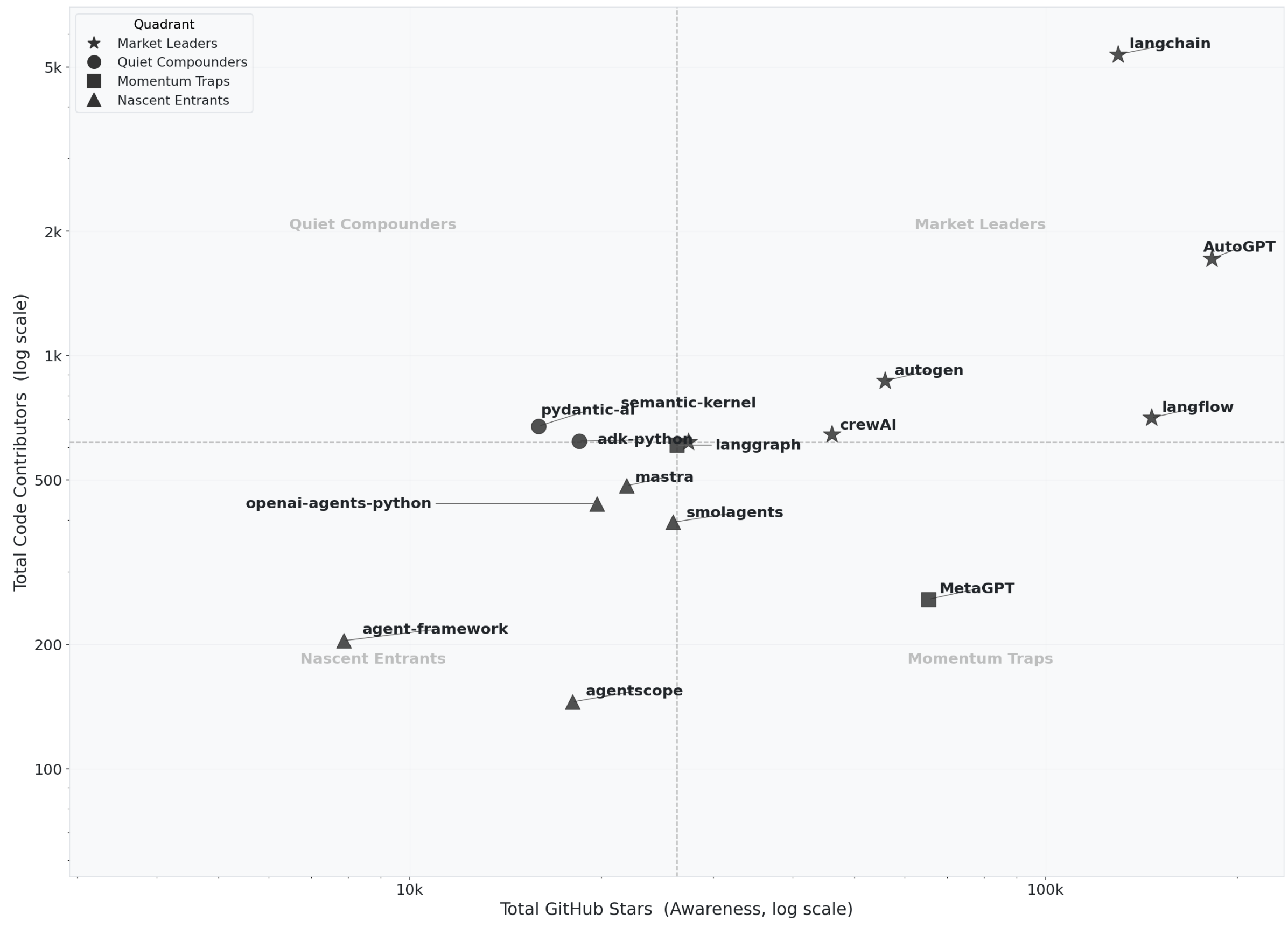

The Momentum Traps quadrant includes projects whose visibility has outpaced contributor adoption. MetaGPT is the clearest example. Despite its high star count of 65,424, its contributor base remains disproportionately small, yielding the lowest contributor density ratio in the dataset at 3.9. Semantic Kernel also falls within this quadrant, though closer to the center. Released in public preview on October 1, 2025, Microsoft Agent Framework merges AutoGen's dynamic multi-agent orchestration with Semantic Kernel's production foundations, likely diverting contributor attention and limiting further independent growth for Semantic Kernel.

The Nascent Entrants quadrant captures projects that are either still early or have yet to convert visibility into broader participation. Within this group, Agent Framework stands out: despite the lowest star count in the dataset, it retains a notably engaged contributor base relative to its visibility. AgentScope, by contrast, has had more time to build momentum but shows the weakest contributor depth in the quadrant. Mastra and openai-agents-python occupy similar positions, with comparable star and contributor counts. Notably, openai-agents-python's modest contributor base despite its OpenAI provenance suggests that institutional backing does not automatically drive community participation. Smolagents approaches the quadrant boundary on awareness, with star counts near the dataset median, though its contributor depth remains below it.

Figure 8 shows absolute positioning, but raw counts do not fully capture the extent to which awareness converts into contribution. To address this, we introduce a contributor density ratio, defined as total code contributors per 1,000 GitHub stars (see Figure 9 and Appendix B). This metric normalizes community depth relative to visibility, allowing frameworks with vastly different star counts to be compared based on how effectively they convert awareness into active contribution. Higher values indicate that a project attracts proportionally more contributors for its level of public attention, a pattern more consistent with utility-driven adoption than with passive interest.

*Figure 9: Contributor Density Ratio*

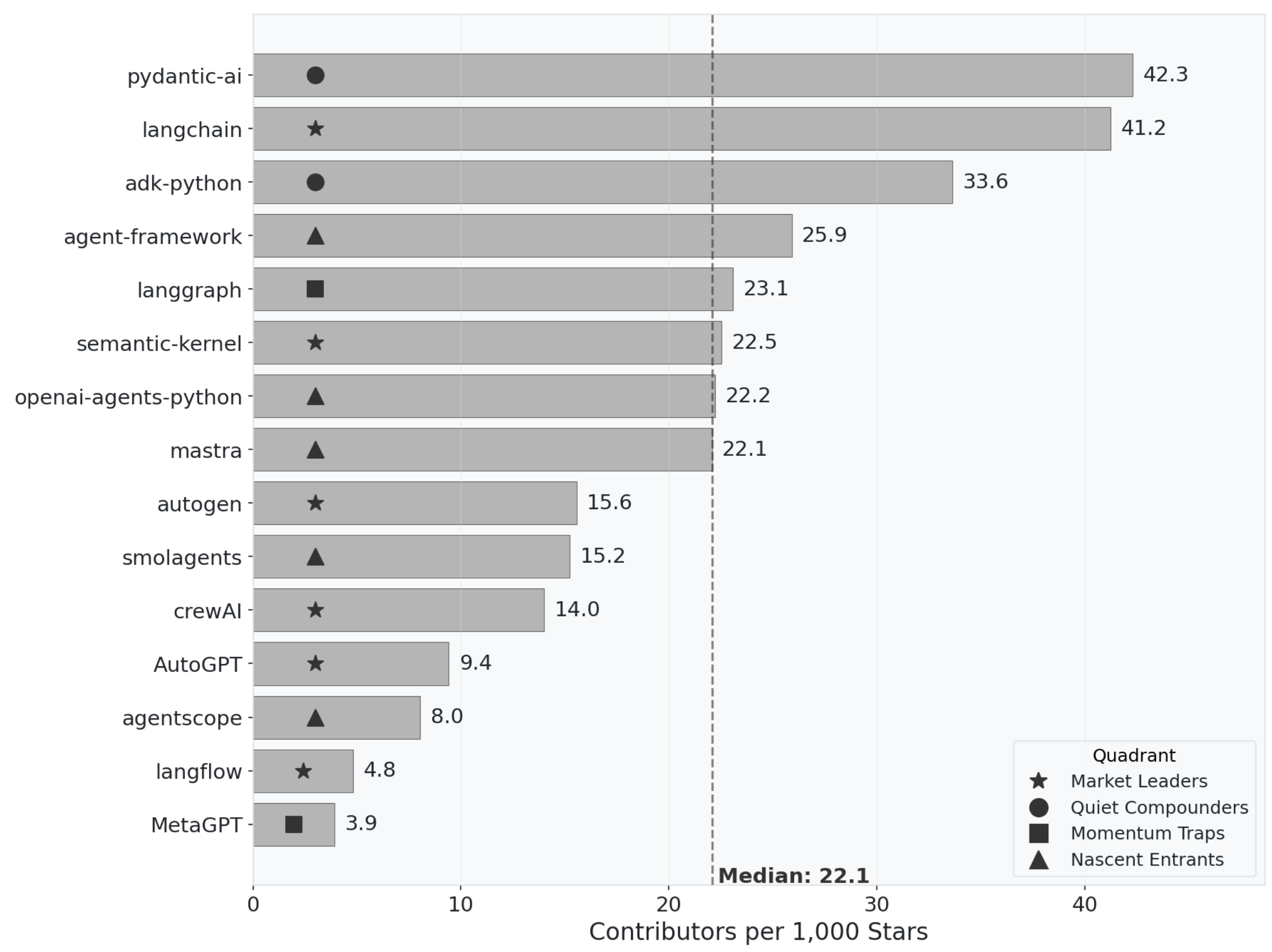

Figure 9 sharpens several patterns from Figure 8. Pydantic-AI leads the dataset at 42.3 contributors per 1,000 stars, followed closely by LangChain at 41.2, confirming that both frameworks convert visibility into contributor participation to a comparable degree. AutoGPT, in contrast, records only 9.4 contributors per 1,000 stars, reinforcing that its high visibility has not translated into equally deep contributor engagement.

The ratio also helps explain why Pydantic-AI and Google ADK stand out as Quiet Compounders. Despite star counts below the dataset median, both frameworks post strong density ratios, Pydantic-AI being first and Google ADK being third at 33.6, indicating that their contributor bases are disproportionately large relative to their visibility.

Several frameworks occupy a broad middle tier, though two sub-clusters are visible within it. LangGraph, Semantic Kernel, OpenAI Agents Python, and Mastra cluster tightly between 22.1 and 23.1 contributors per 1,000 stars, suggesting a comparable degree of contributor conversion despite differences in their origin, backing, and target use cases. AutoGen, smolagents, and CrewAI form a lower cluster between 14.0 and 15.6, a notable result given that all three carry significant institutional or community backing. For AutoGen and CrewAI in particular, which both appear in the Market Leaders quadrant, the relatively modest density ratios suggest that their large absolute contributor counts are driven more by overall visibility than by disproportionate community conversion.

At the other end, MetaGPT records the lowest contributor density ratio in the dataset at 3.9, confirming that its visibility has not translated into a similarly strong community. LangFlow also ranks low at 4.8, further demonstrating that high awareness alone is not a reliable indicator of contributor engagement.

Among lower-visibility repositories, the ratio also distinguishes early-stage projects. Agent Framework posts a ratio of 25.9, suggesting a concentrated but active contributor base relative to its visibility. AgentScope, by contrast, records only 8.0, reinforcing the view that it has struggled to convert even modest visibility into community participation.

Taken together, Figures 8 and 9 provide complementary perspectives on ecosystem health. Figure 8 shows where repositories sit in terms of absolute awareness and contributor engagement, while Figure 9 shows the extent to which awareness is converted into contributor engagement. They help distinguish projects that are merely visible from those that are building deeper and more durable communities.

## 5. Contributor Retention

We examined whether contributors return to make subsequent contributions, defining a contributor as any user with at least one merged pull request or commit. Specifically, we focus on the early contributor cohort, defined as contributors who made their first contribution within the first 180 days of a repository's creation. The steepest decline in retention occurs during the first 30 days, when many contributors do not make a second contribution. Retention rises somewhat at the 60- and 90-day marks for some repositories as contributors have more time to return, but changes little beyond that point. This suggests that contributors who do not re-engage within 90 days are not likely to return later.

We focus on early contributors to make retention behavior more comparable across repositories with different launch dates and cohort structures. Contributors are included if they made their first contribution within the first 180 days of a repository's creation. Repositories with fewer than 30 early contributors, including agentscope and LangGraph with 25 and 13 early contributors respectively, are excluded from the figure to reduce instability from very small sample sizes. In addition, we exclude data points where the observation window extends beyond the period in which the cohort remains intact, so that comparisons are based on consistent cohort sizes over time. We also limit the figure to the first 360 days after the contributors' initial contribution, as retention shows little change beyond that point.

*Figure 10: Cumulative Retention Curves for Early Contributors*

AutoGPT demonstrates a significant and sustained decline in early contributor retention, consistently ranking among the lowest in the dataset throughout the period under review. Within the initial 90 days, over 65% of its contributors had effectively ceased participation and did not return, marking a notable contrast to the project's early enthusiasm and rapid increase in popularity. The retention rate fell to around 30% by Day 30 and remained below 35% even by Day 360. This indicates that although the project initially attracted considerable attention, such interest did not result in sustained user engagement. We argue that AutoGPT's low retention reflects its release prior to the widespread adoption of engineering safeguards that later became standard in agentic frameworks, including loop termination, recursive error handling, and human-in-the-loop checkpoints. Subsequent frameworks such as LangGraph and CrewAI explicitly incorporated these safeguards as core design principles [16][17].

Agent Framework and Semantic Kernel lead the dataset in early cohort retention, reaching 69% and 68% of their founding contributors returning by Day 150 and Day 360 respectively. These cohorts contain 80 and 210 contributors, respectively. Cross-referencing contributor emails and GitHub profile data reveals that 58% of Agent Framework's early contributors and 42% of Semantic Kernel's carry Microsoft affiliation signals (see Appendix C), well above the dataset average of 15%. Agent Framework launched in April 2025 with a concentrated internal Microsoft team. A founding cohort that is more than half Microsoft-affiliated producing 69% retention is more consistent with a managed engineering effort than a spontaneously retained open-source community. The high retention in both repos therefore reflects organizational coordination as much as community stickiness.

Pydantic-AI is an exception worth noting, achieving 63.6% retention from a cohort of 33 contributors. Its relatively high company affiliation rate and strong contributor density ratio suggest a similarly production-oriented contributor base, where returning contributors are likely maintaining integrations with existing production systems rather than engaging out of exploratory interest.

LangChain's retention curve settles at around 45%, mid-table relative to the broader cluster, but its founding cohort of 569 contributors is the largest in the dataset by a considerable margin. Sustaining nearly half of that cohort over a full year is a different achievement from the high retention rates seen at Agent Framework and Semantic Kernel. Unlike those Microsoft-family repos, only 1.2% of LangChain's founding cohort, approximately 8 contributors, show any organizational affiliation with LangChain itself through publicly available GitHub information. Its retention is not explained by employer incentives or coordinated internal teams.

What may better explain it is LangChain's structural position in the ecosystem. The cross-ecosystem analysis shows that LangChain functions as a platform layer that other frameworks build on top of or integrate with. A developer maintaining a production application or another framework built on LangChain has a practical, ongoing reason to keep contributing, independent of any organisational directive. Retention in this context is less about community loyalty and more about dependency: contributors return because the framework is embedded in work they are already doing. That kind of utility-driven engagement is structurally more durable than curiosity-driven contribution, and it may explain why LangChain holds its retention rate across a large and diverse cohort where no single organization is driving the numbers.

The average retention curve stabilizes within the first year, with more than half of these initial contributors failing to return after their initial contribution period. This finding carries practical implications for framework maintainers seeking to grow their contributor communities. Deepening ecosystem integration and supporting downstream framework compatibility may be more effective levers than community engagement initiatives alone. Within this dynamic, the initial 90 days are especially consequential for newcomer retention, factors such as review speed, PR merge rate, and responsiveness to contributor submissions during this window tend to have an outsized influence on whether first-time contributors return [18].

## 6. Conclusions

In summary, this research suggests that the health of open-source AI frameworks cannot be accurately evaluated solely through headline metrics. While star counts are easily accessible and support comparison, they may be influenced by hype cycles, launch effects, and coordinated activities. Metrics that closely reflect contributor behavior, such as cross-ecosystem involvement, migration, and return rates, offer a more dependable assessment of the ecosystem's robustness. Among the 15 frameworks analyzed here, several findings stand out.

First, visible attention and durable adoption often diverge. Some repositories accumulated very high star counts but converted relatively little of that attention into sustained contributor engagement. In this sense, awareness is not irrelevant, but it is incomplete. Star trajectories are better understood as signals of visibility and discovery than as direct measures of ecosystem health. A four-quadrant framework mapping awareness against adoption makes this divergence concrete, distinguishing Market Leaders that convert visibility into contributor depth from Momentum Traps such as MetaGPT and LangFlow where star counts substantially outpace contributor engagement. The finding that openai-agents-python records a modest contributor base despite its OpenAI provenance further suggests that neither viral attention nor institutional backing reliably predicts community depth.

Second, cross-ecosystem contribution patterns suggest that a small number of frameworks function as shared infrastructure within the broader landscape. LangChain consistently appears as a bridge across otherwise unrelated framework families, with 82.5% of cross-ecosystem contributors having engaged with it, indicating a structural role in the ecosystem beyond its raw visibility. A framework can matter not only because many people use it directly, but because it becomes a layer that other frameworks, workflows, and contributors repeatedly pass through. The findings therefore suggest that ecosystem relevance should be understood not only in terms of size, but also in terms of connectivity and architectural centrality.

Third, the Quiet Compounders quadrant reveals that lower-visibility frameworks can show meaningful signs of adoption that aggregate metrics obscure. Pydantic-AI and Google ADK stand out for having relatively strong contributor density compared with their star counts, and both appear to attract contributors engaging for practical, production-oriented reasons rather than exploratory interest. This suggests that projects with modest public visibility may still be developing deeper technical communities, particularly in an ecosystem where experienced contributors may place greater weight on reliability, type safety, and software engineering fit than on popularity alone.

Finally, contributor retention is highly front-loaded. Most repositories experience their steepest drop in returning contributors within the first 30 days, and although retention rises somewhat around the 60- and 90-day marks, it levels off considerably after that. This suggests that the window for converting first-time contributors into repeat participants is relatively short, and the likelihood of re-engagement does not meaningfully increase once that window has passed. AutoGPT illustrates this pattern most clearly, sustaining fewer than 35% of its early contributors across the full observation period despite its exceptional visibility. Factors such as early review speed, responsiveness, and merge outcomes appear to influence whether first-time contributors return, particularly within the initial 90 days, though they do not fully explain retention differences across frameworks. The analysis also finds that high retention does not uniformly signal organic community health. Agent Framework and Semantic

Kernel both achieve retention rates above 65%, but a substantial share of their early contributors carry Microsoft affiliation signals, suggesting that organizational coordination plays a meaningful role alongside community dynamics. LangChain, by contrast, sustains retention across the largest early cohort in the dataset with minimal organizational affiliation, a pattern more consistent with utility-driven engagement: contributors return because the framework is embedded in work they are already doing. More broadly, retention appears to depend not only on contributor experience but also on whether contributors have an ongoing practical reason to return.

Taken together, these results suggest that sustainable adoption correlates less with early visibility than with technical decisions that address practical production needs and with a project's ability to become embedded in how developers work. Frameworks such as LangChain appear to have achieved enduring relevance not simply by attracting attention, but by becoming part of the ecosystem's working infrastructure. More broadly, ecosystem health is better assessed through contributor quality, repeated engagement, and structural relevance than through star counts alone. The frameworks most likely to achieve sustained community adoption may therefore be not those that capture the greatest attention at launch, but those that solve durable technical problems and build communities that continue to return. Future work could extend this analysis by incorporating package download data, API usage metrics, and longitudinal follow-up as newer frameworks such as Google ADK and OpenAI Agents accumulate longer activity histories.

## 7. Limitation and Future Study

This analysis is bounded by its scope and observation window. The 15 repositories were selected to represent the AI agent framework landscape as of March 2026, but the selection favors English-language, GitHub-hosted, open-source projects and excludes commercial products, private repositories, and projects on non-GitHub platforms such as GitLab-hosted or Hugging Face-hosted projects. The sample is also subject to survival bias, since repositories that remained visible and active through early 2026 are more likely to be included than projects that were short-lived, archived early, or never gained traction. This may lead the dataset to overrepresent more durable frameworks and understate failure or abandonment in the broader ecosystem. Repositories launched in late 2024 or 2025, including smolagents, mastra, openai-agents-python, adk-python, and agent-framework, have observation windows of only 11 to 15 months, which limits the strength of retention and adoption conclusions for those projects. In addition, our contributor definition, restricted to pull request and commit authors, captures code contribution but excludes issue authors, reviewers, and other non-code contributors, and may undercount activity conducted through private forks or enterprise mirrors. This is a gap given that enterprise teams frequently maintain private forks of public repositories so production-level engagement may be systematically underrepresented in public GitHub data.

Several methodological choices also involve judgment calls that shape the results. The anomalous star detection approach flags suspicious patterns but cannot definitively separate coordinated manipulation from organic viral spikes. Company affiliation is based on self-reported GitHub profile information and may be incomplete or outdated. The cross-ecosystem and migration analyses identify patterns and associations but do not establish causation. The interpretations throughout the paper are grounded in the data, but confirming causal relationships would require developer surveys, interview studies, or natural experiments such as framework deprecation events.

Future work could extend this analysis in several directions. First, longitudinal tracking at 6- and 12-month intervals would test whether the patterns identified here persist as newer frameworks such as smolagents, Google ADK-Python, and OpenAI Agent SDK Python accumulate longer activity histories. Second, expanding the repository set to include non-Python ecosystems would provide a more complete picture of the broader ecosystem. Third, incorporating package index and manager download data would allow community signals to be compared directly with production usage, addressing the gap between visibility and deployment that this study identifies but cannot fully quantify. Finally, developer surveys targeting contributors in this dataset could validate the interpretations offered here and shed light on the motivations behind cross-ecosystem contribution patterns and framework abandonment.

# Appendix

## Appendix A: Individual Cumulative Star and Monthly Star Histories for All 15 Repositories

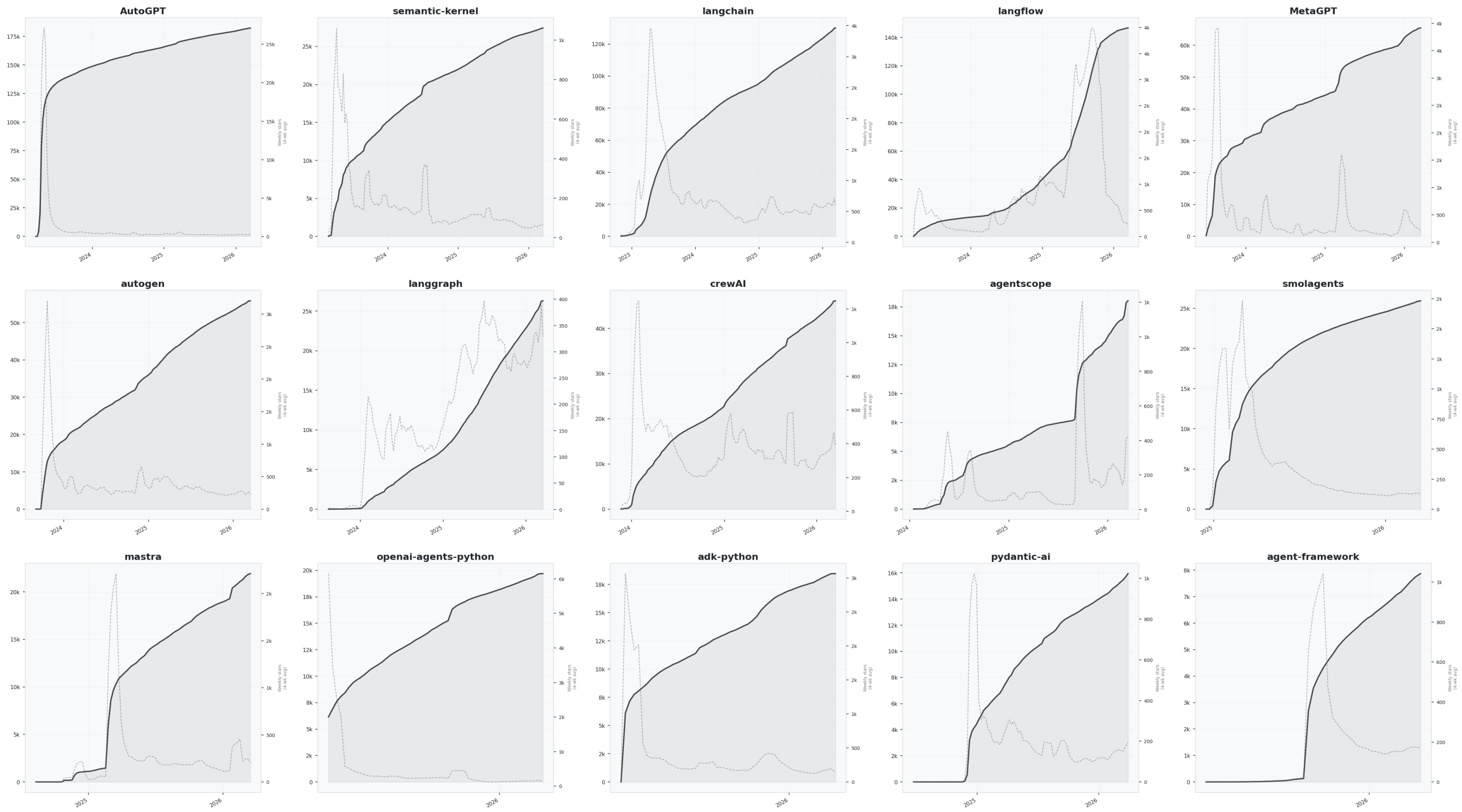

## Appendix B: Repository Awareness and Contributor Engagement Metrics

| Repository Name | Total Stars | Total Contributors | Contributor Density Ratio |
|---|---|---|---|
| pydantic/pydantic-ai | 15950 | 675 | 42.3 |
| langchain-ai/langchain | 130000 | 5362 | 41.2 |
| google/adk-python | 18460 | 621 | 33.6 |
| microsoft/agent-framework | 7872 | 204 | 25.9 |
| langchain-ai/langgraph | 26295 | 607 | 23.1 |
| microsoft/semantic-kernel | 27437 | 618 | 22.5 |
| openai/openai-agents-python | 19699 | 438 | 22.2 |
| mastra-ai/mastra | 21931 | 484 | 22.1 |
| microsoft/autogen | 55842 | 869 | 15.6 |
| huggingface/smolagents | 25940 | 395 | 15.2 |
| crewAIInc/crewAI | 46105 | 645 | 14.0 |
| Significant-Gravitas/AutoGPT | 182405 | 1716 | 9.4 |

| agentscope-ai/agentscope | 18027 | 145 | 8.0 |
|---|---|---|---|
| langflow-ai/langflow | 146655 | 707 | 4.8 |
| FoundationAgents/MetaGPT | 65424 | 257 | 3.9 |

Appendix C: Early Contributor Profile

| Repo | Early Contributors | Median Account Age (yrs) | % Any Company Listed | % Parent Org Affiliated |
|---|---|---|---|---|
| langchain | 569 | 10.7 | 41.7 | 1.2 |
| langflow | 54 | 9.9 | 42.6 | 7.4 |
| AutoGPT | 431 | 9.6 | 32.2 | 1 |
| MetaGPT | 106 | 7 | 30.8 | 0 |
| autogen | 272 | 9.5 | 38.5 | 11.9 |
| crewAI | 99 | 10.9 | 30.6 | 2 |
| semantic-kernel | 210 | 10.4 | 56.9 | 41.6 |
| langgraph | 13 | 11.1 | 33.3 | 25 |
| agentscope | 25 | 7.8 | 36 | 24 |
| pydantic-ai | 33 | 12 | 37 | 14.8 |
| mastra | 28 | 11.8 | 42.9 | 17.9 |
| smolagents | 267 | 9.1 | 34.5 | 6.4 |
| openai-agents-python | 298 | 6.3 | 37.2 | 5.4 |
| adk-python | 338 | 8.2 | 40.9 | 12.4 |
| agent-framework | 80 | 10 | 66.2 | 58.4 |